\documentclass[aps,pre,twocolumn,groupedaddress,showpacs,superscriptaddress,amssymb,amsmath,noeprint]{revtex4-2}
\usepackage{graphicx}
\usepackage{dcolumn}
\usepackage{bm}
\usepackage{hyperref}
\usepackage{cleveref}
\hypersetup{
    colorlinks=true,
    linkcolor=blue,
    urlcolor=blue,
    citecolor=blue
}

\newcommand{\rme}{{\rm e}}
\newcommand{\rmc}{{\rm c}}
\newcommand{\rmd}{{\rm d}}
\newcommand{\rmi}{{\rm i}}

\newcommand{\rmB}{{\rm B}}

\begin{document}

\title{Localization of information driven by stochastic resetting}

\author{Camille Aron}
\email{aron@ens.fr} 
\affiliation{Laboratoire de Physique de l’\'{E}cole Normale Sup\'{e}rieure, ENS, Universit\'{e} PSL,
CNRS, Sorbonne Universit\'{e}, Universit\'{e} Paris Cit\'{e}, F-75005 Paris, France}
\affiliation{Institute of Physics, \'{E}cole Polytechnique F\'{e}d\'{e}rale de Lausanne (EPFL), CH-1015 Lausanne, Switzerland}

\author{Manas Kulkarni}
\email{manas.kulkarni@icts.res.in} 
\affiliation{International Centre for Theoretical Sciences, Tata Institute of Fundamental Research,
Bangalore 560089, India}

\date{\today} 

\begin{abstract}
The dynamics of extended many-body systems are generically chaotic. Classically, a hallmark of chaos is the exponential sensitivity to initial conditions captured by positive Lyapunov exponents.
Supplementing chaotic dynamics with stochastic resetting drives a sharp dynamical phase transition: We show that the Lyapunov spectrum, \textit{i.e.}, the complete set of Lyapunov exponents, abruptly collapses to zero above a critical resetting rate. At criticality, we find a sudden loss of analyticity of the velocity-dependent Lyapunov exponent, which we relate to the transition from ballistic scrambling of information to an arrested regime where information becomes exponentially localized over a characteristic length diverging at criticality with an exponent $\nu = 1/2$ and a dynamical exponent $z=2$.
We illustrate our analytical results on generic chaotic dynamics by numerical simulations of coupled map lattices.
\end{abstract}

\maketitle

\paragraph*{Introduction.}
Chaotic dynamics in extended many-body systems are marked by the rapid scrambling of information, rendering the details of the initial state effectively irretrievable~\cite{d2016quantum,Swingle2018}. Mechanisms such as integrability~\cite{MF2022}, strong disorder~\cite{BK2022}, frequent projective measurements~\cite{Adam2019,fisher2018}, or kinetic constraints~\cite{DRL22,BPM2012} can strongly suppress this scrambling, leading instead to vanishing Lyapunov exponents, exponentially localized correlations, or slow entanglement growth~\cite{LB2017}. Identifying and characterizing such departures from chaos is central to both classical and quantum dynamics, with implications for controlling chaos, protecting quantum information, and engineering novel non-equilibrium states.

Stochastic resetting offers another natural way to mitigate chaos. In its simplest form, resetting consists of returning a dynamical system to its initial configuration at random times~\cite{EM2011,EM2011a,Evans_2014,KMSS2014,Evans_2020,GJ22}. Beyond its applications in first-passage and search problems~\cite{Evans_2013,Mercado-Vasquez_2022}, in many-body systems resetting acts as a non-equilibrium drive that impedes transport, prevents thermalization and can reshape phase diagrams~\cite{Nagar_2023,bressloff2024kuramoto,SG2022,bressloff2024global,GMS2014,BKP19,Sadekar_2020,Karthika_2020,MMS20}.
This has been particularly discussed in the context of the Ising universality class, where resetting was shown to deform the standard Landau-Ginzburg $\varphi^4$ theory by introducing non-analyticities in the potential at $\varphi = 0$, thereby destabilizing the symmetry-broken infrared fixed point, restoring $\mathbb{Z}_2$ invariance, yet producing order-parameter dynamics distinct from those of the conventional thermal paramagnet~\cite{Ak20}.
From the perspective of information dynamics, we recently demonstrated that scrambling, 
can persist at finite resetting rates, but that beyond a critical rate the dynamics undergo a generic transition to a non-chaotic regime, with both the butterfly velocity $v_\mathrm{B}$ and the largest Lyapunov exponent $\lambda_0$ vanishing~\cite{Ak2025}.

In this Letter, we take on the pressing question: What is the precise dynamical nature of this non-equilibrium non-chaotic regime driven by stochastic resetting?
We answer by rigorously deriving how a chaotic Lyapunov spectrum, encoding the asymptotic growth and decay rates of infinitesimal perturbations along different tangent directions in phase space, is renormalized by resetting.
We begin by formulating the generic chaotic many-body dynamics that we have in mind and introduce the coupled logistic map (CLM), which provides a convenient numerical testbed for our analytics. In the absence of resetting, we relate the Lyapunov spectrum $\Lambda(k)$ in the thermodynamic limit to the velocity-dependent Lyapunov exponent (VDLE) $\lambda(v)$, which characterizes the spatiotemporal spread of information. Later, we analyze how stochastic resetting modifies this scrambling of information. Building on this, we derive how $\Lambda(k)$ is renormalized, showing that it collapses abruptly to a flat, vanishing profile at a critical resetting rate. We then argue that this dynamical transition gives rise to a regime where information spreading is arrested in time and exponentially localized in space. Finally, we discuss possible connections with measurement-induced phase transitions.

\label{sec:setup}
\paragraph*{Chaotic dynamics subject to stochastic resetting.}
We consider a field $\phi(x,t)$ evolving from an initial configuration $\phi(x,0)$ under classical chaotic dynamics which are assumed to be local and causal.
This implies a finite Lieb–Robinson velocity $v_{\rm LR}$ beyond which no information can propagate. To simplify, we consider deterministic evolutions where the generator of time evolution (\textit{e.g.}, Hamiltonian flow,  PDE, or discrete non-linear map) is homogeneous in space and time, and symmetric under spatial reflection.
For concreteness, we restrict the discussion to scalar fields in one spatial dimension, though the generalization to higher dimensions is straightforward. 
In addition, the evolution is interspersed with global resetting events at a rate $r$: With probability $r\, \rmd t$, the field is instantaneously returned to its initial state, \textit{i.e.} $\phi(x,t)=\phi(x,0)$ for all $x$.
In this simplest setting, resetting is the only source of fluctuations, but the discussion can be adapted to additional statistical ensembles, such as random initial conditions or noisy dynamics.

\paragraph*{Coupled logistic map (CLM).}
Our analytics are derived in the continuum, but the lessons are also valid in discrete cases. We justify our assumptions, validate our approximations, and illustrate our results by performing numerical simulations of the CLM which is an archetypal discrete non-linear map. The choice of the CLM is guided by its relative simplicity to simulate and because its chaotic properties in the absence of resetting have already been studied extensively~\cite{Kaneko92,KANEKO198960,chazottes2005dynamics,hagerstrom2012experimental,losson1996,MS2022,Notenson2023,Politi_book_2016}.
It is a one-dimensional chain of $L$ sites, each hosting a logistic map that is diffusively coupled to its two nearest neighbors. The initial state at $t=0$ is specified by $\phi_{x,0}$ drawn randomly in $[0,1]$ for $x=1,2, \ldots, L$. Note that all the results reported below are obtained from a \emph{single} such realization.
When subject to resetting, the dynamics are generated by the stochastic nonlinear map $\phi_{x,\,t} \to \phi_{x,\, t+1}$ with
\begin{align} \label{eq:CLM}
\phi_{x,\, t+1} \!= \!
\begin{cases}
    (1-c) f(\phi_{x,\,t})\! +\! \frac{c}{2} \! \left[ f(\phi_{x-1,\,t}) \!+\! f(\phi_{x+1,\,t}) \right] \\ \qquad \mbox{ with probability } 1-r \\
\phi_{x,\, 0}  \mbox{ with probability } r,
\end{cases}
\hspace{-1em}
\end{align}
together with periodic boundary conditions. Here, $f(y) = a \, y(1-y)$ is the logistic function with the parameter $a \in [0,4]$ and $0 \leq c \leq 1$ is a diffusion parameter. $0 \leq \phi_{x,\,t} \leq 1$ for all $x$ and $t \geq 0$.
In practice, we work with $a=4$ and $c=0.1$ for which we have carefully checked that the deterministic dynamics (at $r=0$) are chaotic and reach a stationary state that is uniform in space and time.
We avoid boundary effects by working with $v_{\rm LR} \, t < L/2$ and, here, $v_{\rm LR} = 1$. 
Note that, due to the discreteness of time in the CLM, continuous-time expressions such as $\exp(-r \, \rmd t)$ should be replaced with $1-r$.
For methodological details on simulating coupled maps under resetting, see Ref.~\cite{Ak2025}.

\label{sec:Osedelets}
\paragraph*{Oseledets' theorem.}
We begin by characterizing the chaotic dynamics in the absence of resetting, \textit{i.e.}, at $r=0$.
The Lyapunov spectrum $\Lambda(k)$ is related, via Oseledets’ multiplicative ergodic theorem~\cite{oseledets1968,Raghunathan1979}, to a classical out-of-time-ordered correlator (OTOC):
\begin{align} \label{eq:Oseledets}
\Lambda(k) = \lim_{t \to \infty} \frac{1}{t} \log |D(k;t)|,
\end{align}
where $k$ labels the instantaneous eigenvalues $D(k;t)$ of the OTOC defined as (at $r=0$)
\begin{align} \label{eq:OTOC}
D(x,x';t) := \left| \frac{\delta \phi(x,t)}{\delta \phi(x',0)} \right|.
\end{align}
$D(x,x';t)$ quantifies how an infinitesimal local perturbation of the initial configuration propagates across space and time~\cite{2018-das--bhattacharjee,2018-bilitweski--moessner}. The terminology ``out-of-time-ordered correlator'' originates from the quantum setting, where such quantities typically appear as squared commutators of spatially separated operators, evaluated at non-monotonic sequences of times~\cite{larkin1969,S2014,MSS16,RS16}. 
More generally, the definition of OTOCs may be supplemented by ensemble averages over initial conditions or stochastic forces. In Eq.~(\ref{eq:OTOC}), we use the absolute-value norm, though one may instead consider the Euclidean norm or other norms appropriate to the problem at hand.

\begin{figure} 
    \centering
    \includegraphics[width=0.95\linewidth]{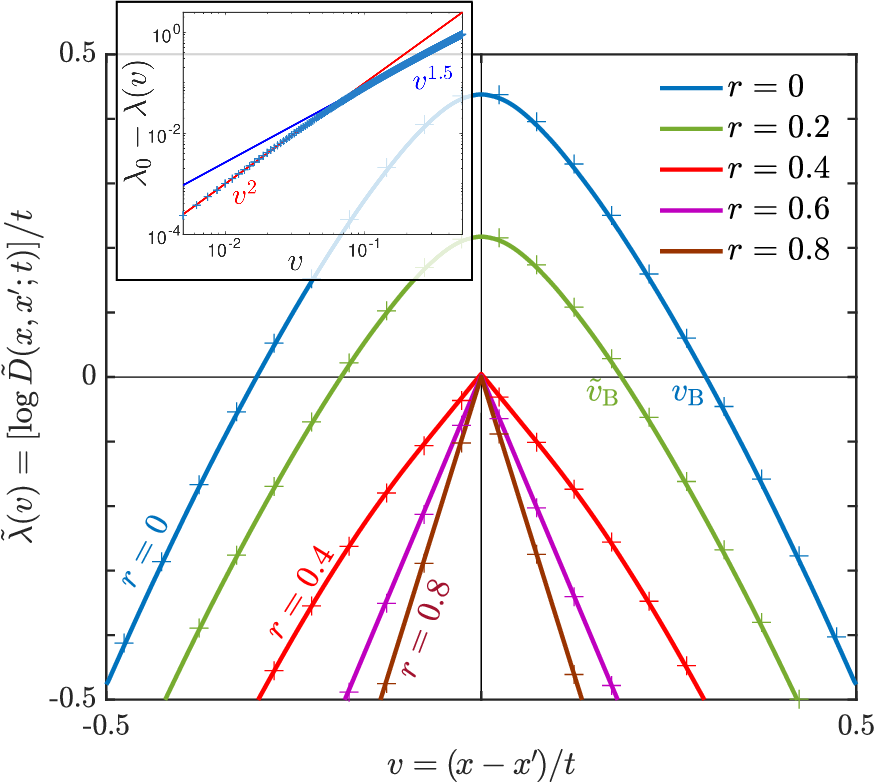}
    \caption{
    Velocity-dependent Lyapunov exponent (VDLE), $\tilde \lambda(v)$, in the CLM model [Eq.~(\ref{eq:CLM})] with $a = 4$, $c = 0.1$, for increasing resetting rates $r$ indicated in the key. The VDLE ansatzes in Eq.~(\ref{eq:VDLE_ansatz}) and below Eq.~(\ref{eq:renewal_cont_lattice}) are validated by comparing data for $L = 401$, $t=200$ (lines) with $L=801$, $t=400$ (crosses).
    When $r \leq r_\rmc \approx  0.37(2)$, $\tilde \lambda(v) = \lambda(v) + \log(1-r)$ as predicted in Eq.~(\ref{eq:lambda_v_shift}). When  $r > r_\rmc$, $\tilde \lambda(v)$ develops a non-analytic cusp at $v=0$, with linear branches predicted in Eq.~(\ref{eq:linear_branch}). 
   Inset: log-log plot of the $r=0$ data, consistent with the small-velocity expansion in Eq.~(\ref{eq:expansion}).
    }   \label{fig:VDLE}
\end{figure}

\paragraph*{Velocity-dependent Lyapunov exponent (VDLE).}
At late times, and in the thermodynamic limit, we assume that the spatiotemporal growth of the OTOC is translation and reflection symmetric~\footnote{If starting from an initial configuration that breaks translation or reflection symmetries, we assume that both symmetries are restored by the dynamics at late times, or that an appropriate ensemble average statistically restores them.}, and described by the ansatz (at $r=0$)
\begin{align} \label{eq:VDLE_ansatz}
    D(x,x';t) \sim \exp\!\left[ \lambda\!\left(\tfrac{x-x'}{t}\right) t \right],
\end{align}
for $|x-x'| < v_{\rm LR} t$, while $D(x,x';t) = 0$ outside the causal light cone.  
Here, $\lambda(v)$ is the VDLE, first introduced as the comoving~\cite{DEISSLER1987397,KANEKO1986436} or convective Lyapunov exponent~\cite{Politi_book_2016}.
This ansatz is known to capture the phenomenology of a wide class of classical and semiclassical chaotic systems~\cite{DEISSLER1987397,2018-khemani--nahum,2020-chatterjee--kulkarni,Moessner2021,RHK2023}.  
Reflection symmetry implies that $\lambda(v)$ is an even function of $v$. We assume it is maximal at $v=0$ with $\lambda_0 := \lambda(0) > 0$ coinciding with the largest Lyapunov exponent, and monotonically decreasing for $v>0$.
In systems with local dynamics, $\lambda(v)$ is generally expected to be analytic around $v=0$. Therefore, we parametrize its small-velocity expansion as
\begin{align} \label{eq:expansion}
    \lambda(v \simeq 0) =  \lambda_0 \, [1 - (v/v_0)^2] + O(v^4),
\end{align}
with $v_0 > 0$.
The function $\lambda(v)$ vanishes at a finite butterfly velocity $v_{\rm B} := \lambda^{-1}(0)$, which sets the maximum propagation speed of infinitesimal perturbations. The late-time dynamics feature a ballistic wavefront propagating at $v_{\rm B}$, with the exponential growth rate at the front given by $\lambda_{\rm B} = -v_{\rm B}\, \lambda'(v_{\rm B}) > 0$.
Except for the mild assumptions above, our results do not require further specifying the functional form of $\lambda(v)$.
In the App.~\ref{app:parabolic}, we detail the special case of a parabolic VDLE of the form $\lambda(v) = \lambda_0 [1-(v/v_{\rm B})^2]$ that can be treated analytically without any approximation.
In Fig.~\ref{fig:VDLE}, we numerically validate the ansatz in Eq.~(\ref{eq:VDLE_ansatz}) and extract $\lambda(v)$ in the specific case of the CLM where the data are consistent with Eq.~(\ref{eq:expansion}).
See also Ref.~\cite{KANEKO1986436}. 

\paragraph*{Lyapunov spectrum.}
Using the late-time translational invariance, the eigenvalues of $D(x-x';t)$  in the thermodynamic limit are simply obtained as its Fourier coefficients, and the Lyapunov spectrum can be expressed as
\begin{align} \label{eq:spectrum_from_VDLE}
        \Lambda(k) = \lim\limits_{t\to\infty} \frac{1}{t} \log \left| 
    \int \rmd v \, \exp\left[\left(\lambda(v)-\rmi k v\right) t\right]
    \right|,
\end{align}
where $k \in \mathbb{R}$ is now identified with the momentum in reciprocal space.
The existence of a well-defined thermodynamic limit of the Lyapunov spectrum was first conjectured by Ruelle~\cite{Ruelle1982} and later confirmed in a variety of settings, including Hamiltonian flows such as the Fermi–Pasta–Ulam–Tsingou chain~\cite{Livi_1986}, PDEs such as  the Kuramoto–Sivashinsky or the complex Ginzburg–Landau equations~\cite{Chate2009,Chate2011}, and coupled map lattices such as the CLM~\cite{Politi_book_2016}. Here, the VDLE ansatz in Eq~\eqref{eq:VDLE_ansatz}, by construction, ensures the existence of this thermodynamic limit.

The leading Lyapunov exponents are obtained by performing a small-$k$ expansion of Eq.~(\ref{eq:spectrum_from_VDLE}). This relies on the small-velocity expansion in Eq.~(\ref{eq:expansion}) together with a saddle-point evaluation, controlled by the large-$t$ limit. See the details of the computation in App.~\ref{app:lambda_k}. We obtain
\begin{align} \label{eq:spectrum}
\Lambda(k \simeq 0) &= \lambda_0 - \lambda_0 \left( \frac{v_0 k}{2\lambda_0} \right)^{2} + \ldots,
\end{align}
where the ellipsis denotes higher-order terms in $k$ arising from subleading contributions in Eq.~(\ref{eq:expansion}).

The result in Eq.~(\ref{eq:spectrum}) is remarkable in that the density of Lyapunov exponents near the upper edge of the spectrum exhibits a square-root singularity, going as $1/(\lambda_0 - \Lambda)^{1/2}$, entirely determined by the behavior of the VDLE $\lambda(v)$ in the vicinity of $v=0$. Importantly, this relation holds for essentially any extended chaotic many-body system in the thermodynamic limit where the VDLE ansatz of Eqs.~(\ref{eq:VDLE_ansatz}) and~(\ref{eq:expansion}) applies, providing a powerful analytical handle given the numerical difficulty of accessing full Lyapunov spectra in large systems. In the special case of a parabolic VDLE, one obtains a parabolic spectrum: $\Lambda(k) = \lambda_0 - k^2/|2\lambda''(0)|$ for all $k$.

Noteworthy enough, for a \emph{discrete} one-dimensional chain, where the Brillouin zone is restricted to $k \in [-\pi, \pi)$, one recovers the same expression as in Eq.~(\ref{eq:spectrum}), see App.~\ref{app:discrete} for details.

\paragraph*{Scrambling under resetting.}
\label{sec:VDLE}
Let us now turn on the stochastic resetting, \textit{i.e.}, $r>0$. The OTOC is defined as
\begin{align} \label{eq:OTOC_resetting}
    \tilde D(x,x';t) := \left\langle \left| \frac{\delta \phi(x,t)}{\delta \phi(x',0)} \right|\right\rangle_r ,
\end{align}
where $\langle \ldots \rangle_r$ is the average over the ensemble of all possible realizations of the resetting times.
The tilde is used here and below to denote quantities that are renormalized by resetting.
Owing to the simplicity of the stochastic resetting protocol, the OTOC in the presence of resetting is related to the OTOC without resetting via a ``renewal equation'' that reads
\begin{align} \label{eq:renewal_cont_lattice}
\tilde D(x,x';t) \!=\!r \! \! \int_0^t \!\! \rmd \tau \, \rme^{-r \tau} D(x,x';\tau) + \rme^{-r t} D(x,x';t).
\end{align}
At late times, we find that $\tilde D(x,x';t) \sim \exp[ \tilde \lambda (\frac{x-x'}{t}) t ]$ where the functional form of the renormalized VDLE $\tilde \lambda(v)$ crucially depends on the relative distance to a critical resetting rate set by the largest Lyapunov exponent in the absence of resetting, $r_\rmc = \lambda_0$.
The details are provided in App.~\ref{app:lambda_v}.
Let us emphasize that $\tilde \lambda(v)$ does not directly relate to the standard definition of Lyapunov exponents since the average over trajectories is performed at the level of the OTOC rather than to its logarithm.
While averaging over resetting histories (or any other statistical ensemble) erases the information about individual trajectories, the notion of exponential sensitivity to initial conditions remains meaningful and can still be probed with OTOCs. This is notably the point of view taken in semiclassical approaches to quantum chaos.
For $r \leq  r_\rmc$, the VDLE is shifted down by the resetting rate,
\begin{align} \label{eq:lambda_v_shift}
    \tilde \lambda(v) = \lambda(v) - r  \quad \mbox{if } r \leq  r_\rmc.
\end{align}
The simplicity of this shift is rooted in the competing exponentials: Lyapunov growth and the Poisson process.
In particular, the largest Lyapunov exponent under resetting $\lambda_0 -r$ is positive: While the exponential sensitivity to initial conditions is reduced by a finite resetting rate, the dynamics are still chaotic.
As the critical point is approached from below, $r\to r_\rmc^-$, the renormalized butterfly velocity vanishes as $\tilde v_\rmB \sim (r_\rmc - r)^{1/2}$.
See Ref.~\cite{Ak2025} for a dynamical characterization of this chaotic regime.
For $r > r_\rmc$, we find a more involved reduction of the VDLE, with
\begin{align} \label{eq:linear_branch}
    \tilde \lambda(v) = 
    \left\{
    \begin{array}{ll}
      \lambda'(v^*) \, |v|   & \mbox{ for } |v| < v^*  \\
     \lambda(v) - r    & \mbox{ for } |v| \geq v^*
    \end{array}
    \right.
     \quad \mbox{if } r >  r_\rmc,
\end{align}
where $v^* > 0$ is the unique solution of
\begin{align}
 \lambda'(v^*) = \frac{\lambda(v^*)-r}{v^*} < 0.
\end{align}
The effective VDLE is now everywhere non-positive: $\tilde \lambda(v) \leq 0$. This indicates that the dynamics are no longer chaotic and $\tilde v_\rmB =0$.
Strikingly, the maximum of $\tilde \lambda(v)$ located at $v=0$ is pinned to $\tilde \lambda_0 = 0$ for all $r>r_\rmc$, while $\tilde \lambda(v)$ develops linear branches extending from $v=0$ to $\pm v^*$, where it remains continuous but is non-analytic. The findings in Eq.~(\ref{eq:linear_branch}) are supported in Fig.~\ref{fig:VDLE} in the context of the CLM. It compellingly shows that strong resetting induces a sudden loss of analyticity in $\tilde \lambda(v)$ at $v=0$ for $r > r_\rmc$, changing from a smooth parabolic behavior to a cusp that sharpens as $r$ increases further.

\begin{figure}   
    \centering
    \includegraphics[width=0.95\linewidth]{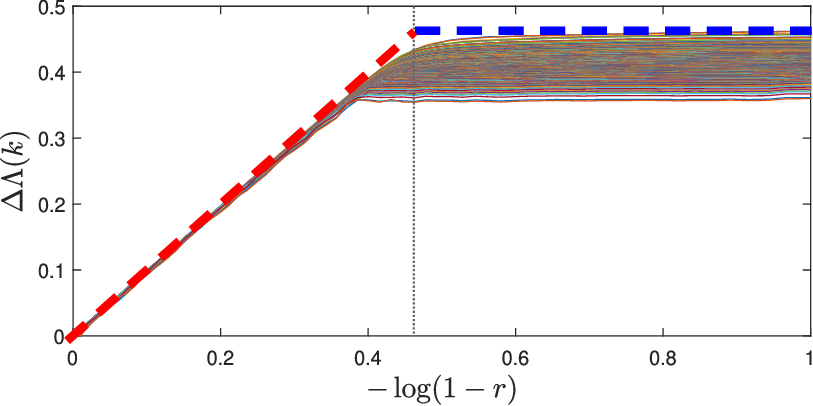}
    \caption{
    Variation of the Lyapunov spectrum across the dynamical phase transition driven by stochastic resetting. The analytical results in Eq.~(\ref{eq:spectrum_r}) are tested in the CLM model [Eq.~(\ref{eq:CLM})].
    For resetting rates $r < r_\rmc \approx 0.37(2)$, $\Delta \Lambda(k) := \Lambda(k) - \tilde \Lambda(k)$ is independent of $k$ for all $k$ and falls on $-\log(1-r)$ shown in dashed red.
    In contrast, for $r > r_\rmc$, the collapse of $\tilde \Lambda(k)$ manifests itself by the dependence of $\Delta \Lambda(k)$ on $k$ and its independence on $r$.
    The blue dashed line is a horizontal guide for the eye. 
    The spectra are computed for $L=401$ at time $t=200$, with a temporal average over $\Delta t = 10$. The transition between the two regimes sharpens when simultaneously increasing $L$ and $t$.
    The other parameters are as in Fig.~\ref{fig:VDLE}.
    } \label{fig:spectrum}
\end{figure}

\paragraph*{Lyapunov spectrum under resetting.} 
\label{sec:spectrum}
We define the Lyapunov spectrum in the presence of resetting, $\tilde \Lambda(k)$, through Eq.~(\ref{eq:Oseledets}) where $D$ is now replaced by $\tilde D$ defined in~(\ref{eq:OTOC_resetting}).
We refer the reader to App.~\ref{app:lambda_k} for the details of the computations.
We find a spectral phase transition,
\begin{align} \label{eq:spectrum_r}
\tilde \Lambda(k) =
\begin{cases}
 \Lambda(k) - r   & \mbox{if } r < r_\rmc = \lambda_0\\
         0 & \mbox{if }  r \geq r_\rmc,
\end{cases}
\end{align}
for all $k$.
Below criticality, the Lyapunov spectrum is shifted down by the resetting rate. In particular, the density of Lyapunov exponents at the upper edge of the spectrum is still governed by a Van-Hove singularity in $1/(\lambda_0-r-\tilde \Lambda)^{1/2}$. At criticality and above, the spectrum is found to collapse to a flat, zero spectrum. This is the central result of our work, describing an abrupt transition from an extended distribution of Lyapunov exponents, peaked at $\lambda_0 - r$, to a distribution where all Lyapunov exponents are maximally degenerate and vanishing.

We illustrate this spectral transition in Fig.~\ref{fig:spectrum} with the example of the CLM. Even for discrete spacetimes, computing the Lyapunov spectrum is a numerically challenging task. Indeed, the standard QR decomposition methods~\cite{Politi_book_2016} cannot be readily used in the resetting context where the notion of individual trajectories is lost after averaging over the resetting times. 
We therefore resort to brute-force computations based on Oseledets’ approach: $\tilde \Lambda(k)$ is extracted from the eigenvalues of $\tilde D(x,x';t)$, obtained via the renewal equation from $D(x,t';t)$, itself computed from the tangent dynamics of a single deterministic trajectory at $r=0$. This procedure is known to be quite sensitive to the growth of numerical errors. In practice, we mitigate those errors by computing the different spectra at a finite time $t = 200$ rather than trying to numerically converge the late-time spectra.
The predictions in Eq.~(\ref{eq:spectrum_r}) are tested by plotting the relative distance between the spectra in the absence and in the presence of resetting, $\Delta \Lambda(k):=  \Lambda(k) - \tilde \Lambda(k)$, as a function of the resetting rate. 
In the regime $r<r_\rmc$, the numerically extracted $\Delta \Lambda(k)$ clearly does not depend on $k$ and falls on $-\log(1-r)$ for all $k$.
In the regime $r>r_\rmc$, the dependence of $\Lambda(k)$ on $k$ and its independence on $r$ clearly manifest themselves as a thick bundle of horizontal lines. Altogether, we find a very good agreement between $\Delta \Lambda(k)$ computed numerically and the analytical prediction made in Eq.~(\ref{eq:spectrum_r}).

\paragraph*{Localization regime.}
\label{sec:localization}
We now discuss how the flat vanishing Lyapunov spectrum for $r > r_\rmc$ materializes in terms of the spatiotemporal spreading of information and correlations.
Equation~(\ref{eq:linear_branch}) implies that the late-time limit of the OTOC $\tilde D(x,x';t)$ saturates to a spatial profile that is exponentially localized around the perturbation $x'$,
\begin{align} \label{eq:local1}
\lim\limits_{t\to\infty} \tilde D(x,x';t) \sim \exp\left[ -|x-x'| / \xi_r \right],
\end{align}
with the length scale
\begin{align} \label{eq:local2}
 \xi_r := -\frac{1}{\lambda'(v^*)} \stackrel{r\to r_\rmc^+}{\longrightarrow} \frac{1}{2} \frac{v_0}{\lambda_0} \left( \frac{\lambda_0}{r-r_\rmc}\right)^{1/2}.
\end{align}
The approach to the steady-state is controlled by the timescale $\tau_r := 1/(r - r_\rmc)$.
Equations~(\ref{eq:local1}) and~(\ref{eq:local2}) demonstrate that the spatiotemporal spread of information is brought to a complete halt in time, with a residual spatial smearing occurring on a length scale $\xi_r$ which diverges in approaching criticality from above with an exponent $\nu = 1/2$ and a dynamical exponent $z=2$.
Note that Eq.~(\ref{eq:local2}) can be generalized to cases in which the small-velocity expansion in Eq.~(\ref{eq:expansion}) starts at order $\alpha > 2$, \textit{i.e.} $\lambda(v\simeq 0) \simeq \lambda_0(1-|v/v_0|^\alpha)$, yielding the critical $\nu = (\alpha-1)/\alpha$.

We now turn to standard correlations, whose behavior under resetting is simpler, if not trivial. In extended many-body systems without conserved quantities and away from phase transitions, one expects equal-time connected two-point functions $C(x,x';t):= \langle \phi(x,t) \phi(x',t) \rangle_{\rm c}$ to decay exponentially in space at large distances. 
Under stochastic resetting, this exponential relaxation is preserved by the renewal structure, leading to a steady-state correlation
\begin{align}
\lim_{t\to\infty} \tilde C(x,x';t) &= r \int_0^\infty \rmd \tau \, \rme^{-r \tau} \, C(x,x';\tau) \\
&\sim \exp[-|x-x'|/\xi_{\rm R}].
\end{align}
where $\xi_{\rm R}$ is a nonuniversal relaxation length scale.
Overall, both standard correlations and the OTOC exhibit exponential localization in space and fail to relax in time, two hallmarks of localization phenomena.

\paragraph*{Conclusions and Outlook.} 
\label{sec:discussion}
Using Oseledets’ connection between OTOCs and Lyapunov spectra, we demonstrated that stochastic resetting drives a dynamical phase transition from an information-scrambling to an information-localized regime, marked by the onset of non-analyticity in the renormalized VDLE $\tilde \lambda(v)$. Strikingly, both the singular upper edge of the density of Lyapunov exponents in the scrambling phase and the critical exponent of the localization length $\xi_r$ in the localized phase are governed by the small-velocity asymptotics of the original VDLE $\lambda(v)$. This contrasts with the ballistic butterfly front, which is instead governed by $\tilde \lambda(v)$ near $\tilde v_\rmB$.

At first sight, the information-localized regime for $r \geq r_\rmc$ resembles integrable dynamics: Vanishing Lyapunov exponents lead to a collapsed spectrum, much as in systems constrained by an extensive number of conservation laws.
Yet, the more compelling connection is with measurement-induced phase transitions (MIPTs) in quantum chaotic systems~\cite{Adam2019,Potter2022,AdamReview2023,Schiro2022}. Classical constructions—via large-$N$ limits, inter-replica dynamics, or learnability transitions in monitored stochastic processes~\cite{Moessner2021,Moessner2022,Vasseur2022,Vasseur2025}—have so far failed to reproduce the robust dynamical exponent $z=1$ signaling emergent spacetime symmetry in MIPTs. Resetting dynamics, with its abrupt projection onto a restricted subset of configurations, naturally echoes features of the Born rule and can even generate long-range correlations~\cite{BLMS23,BKMS2024,SC2025,Kulkarni_2025} reminiscent of quantum measurement protocols such as teleportation.
This perspective suggests an interesting direction: Probing the stability of the information-localized phase under partial resetting~\cite{shamikpre,Shamik2025}, where only a fraction of the degrees of freedom are reset.

\medskip

\paragraph*{Acknowledgements.}
We thank Romain Vasseur and Shamik Gupta for insightful discussions.
M.K. acknowledges support from the Department of Atomic Energy, Government of India, under Project No. RTI4001. M.K. also thanks the VAJRA faculty scheme (No. VJR/2019/000079) from the Science and Engineering Research Board (SERB), Department of Science and Technology, Government of India. 
M.K. thanks the Institute of Physics at EPFL in Lausanne for their hospitality.


\setcounter{equation}{0}
\setcounter{figure}{0}
\renewcommand{\theequation}{A\arabic{equation}}
\renewcommand{\thefigure}{A\arabic{figure}}

\appendix
\section{Velocity-dependent Lyapunov exponent under resetting}
\label{app:lambda_v}
The objective is to characterize the late-time spatiotemporal spread of information in a chaotic extended many-body system subject to stochastic resetting ($r > 0$).
We show that at late times
\begin{align}
   \tilde D(x,x';t) \sim \exp\left[ \tilde \lambda \left(\frac{x-x'}{t}\right) t \right],
\end{align}
and we compute the renormalized VDLE $\tilde \lambda(v)$ in terms of the original VDLE $\lambda(v)$ in the absence of resetting ($r=0$).

We start with the assumption that the late-time OTOC in the absence of resetting ($r=0$) can be described by a VDLE $\lambda(v)$, \textit{i.e.}
\begin{align} \label{eq:ansatz}
D(x,x';t) \sim \exp\left[ \lambda \left(\frac{x-x'}{t}\right) t \right],
\end{align}
where $\lambda(v)$ is assumed to be a continuous and even function of $v$ that monotonously decreases from $\lambda_0 := \lambda(v=0) > 0$.
Working at late times, let us postulate that we can use the ansatz (\ref{eq:ansatz}) in the renewal equation~given in Eq.~(9) of the main text. This assumption will be checked \textit{a posteriori}. Doing so, we obtain 
\begin{align} 
\tilde D(x;t) \sim r \int_0^t \!\! \rmd \tau \, \rme^{[\lambda(x/\tau) -r] \tau}+ \rme^{[\lambda(x/t)-r] t}.
\end{align}
Let us work on the light ray parameterized by $x = vt$ with $v\geq0$, we have
\begin{align} \label{eq:D_v}
\tilde D(v;t)  \sim r \int_0^t \!\! \rmd \tau \, \rme^{[\lambda(vt/\tau) -r] \tau}+ \rme^{[\lambda(v)-r] t}.
\end{align}
The first term above can be approximated using a saddle-point approximation (Laplace's method). The integrand is maximal at the time $\tau^*$ that solves
\begin{align}
0 =     \partial_\tau \left([\lambda(vt/\tau) -r ]\tau \right).
\end{align}
Using $v^* := v t/\tau^*$, the location of the saddle is rewritten as
\begin{align} \label{eq:v_star}
    \lambda(v^*) - r = v^* \lambda'(v^*),
\end{align}
Importantly, the existence of the saddle point is conditioned to $\tau^* < t$ or, equivalently, $v < v^*$. Otherwise, the integrand in Eq.~(\ref{eq:D_v}) is simply maximal at its bound $\tau = t$ and we obtain $\tilde D(v > v^*;t)  \sim  \rme^{[\lambda(v)-r] t} $.
A careful inspection of Eq.~(\ref{eq:v_star}), using $\lambda'(v) < 0$, shows that the existence of a 
saddle point $v^*$ also requires $r \geq r_\rmc$ with the critical resetting rate $r_\rmc = \lambda_0$. 
In this case, the renormalized OTOC is approximated as the sum of two competing exponentials
\begin{align} \label{eq:D_intermediate}
    \tilde D(v < v^*;t)  \sim A \, r  \, \rme^{\lambda'(v^*) vt}+ \rme^{[\lambda(v)-r] t},
\end{align}
where $A$ denotes inconsequential (subexponential) contributions.
Using the property (stemming from the saddle-point condition)
\begin{align}
\lambda'(v^*)  \geq \frac{\lambda(v) - r}{v},
\end{align}
the first term in Eq.~(\ref{eq:D_intermediate}) dominates.
In the case  $r \leq r_\rmc$, there is no saddle point and the integrand in Eq.~(\ref{eq:D_v}) is simply maximal at its bound $\tau = t$ and $\tilde D(v;t)  \sim  \rme^{[\lambda(v)-r] t} $.
In all cases, the location of the maximum scales linearly with $t$. This justifies, a posteriori, the use of the bare velocity-dependent Lyapunov ansatz inside the renewal equation: The early-time behavior of the bare OTOC is irrelevant to the renormalized OTOC at late times.
Altogether, restoring the possibility of $v<0$, we obtain the renomalized VDLE
\begin{align}
\tilde\lambda(v) =
\begin{cases}
\lambda(v) - r & \mbox{if } r \leq r_\rmc    \\
\begin{cases}
 \lambda'(v^*) \, |v|  &\mbox{for } |v| \leq v^* \\
 \lambda(v) - r & \mbox{for } |v| \geq v^* 
\end{cases}
& \mbox{if } r > r_\rmc,
\end{cases}
\end{align}
where $v^*$ is determined by Eq.~(\ref{eq:v_star}).
Importantly, in the case  $r > r_\rmc$, the renormalized VDLE develops linear branches for $|v| < v^*$. 
Note that $\tilde \lambda(v)$ is continuous but not analytic at $v=0$ and $v=\pm v^*$.
As one approaches the critical resetting rate from above, the linear branch of the renormalized VDLE is controlled by the small-velocity expansion of $\lambda(v)$ 
\begin{align} 
 \lambda(v \simeq 0) =  \lambda_0 \, [1 - (v/v_0)^2] + O(v^4),
\end{align}
where $v_0 > 0$, and it shrinks as
\begin{align}
    v^* \stackrel{r\to r_\rmc^+}{\longrightarrow} 
    v_0 \left(\frac{r-r_\rmc}{\lambda_0}\right)^{1/2}.
\end{align}

Altogether, the largest Lyapunov rate is always located at $v=0$ and obeys
\begin{align}
    \tilde \lambda_0 = \max(\lambda_0 - r, 0).
\end{align}

\section{Lyapunov spectrum}
\label{app:lambda_k}
The objectives are twofold: (i) to compute the Lyapunov spectrum $\Lambda(k)$ of a chaotic extended many-body system, assuming that the late-time growth and spatial spreading of the OTOC are captured by
\begin{align}
D(x,x';t) \sim \exp\left[\lambda \left(\frac{x-x'}{t}\right) t \right],
\end{align}
where the VDLE $\lambda(v)$ is continuous, even in $v$, and monotonically decreases from its maximum $\lambda_0 := \lambda(0) \ge 0$; and (ii) to apply this formalism to chaotic dynamics subject to stochastic resetting.


\paragraph*{(i) From $\lambda(v)$ to $\Lambda(k)$.}
Noting that $ D(x,x';t)$ is a real symmetric square matrix in its spatial indices, this ensures that its eigenvalues are real. Furthermore, the translational invariance ensures that the standard momentum basis diagonalizes the matrix.
Therefore, the Oseledets' expression of the Lyapunov spectrum in Eq.~(2) of the main text simplifies as
\begin{align}
     \Lambda(k) = \lim\limits_{t\to\infty} \frac{1}{t} \log \left| 
    \int \rmd x \, \rme^{-\rmi k x} \exp\left[ \lambda \left(\frac{x}{t}\right) t \right]
    \right|,
\end{align}
where the momentum $k\in\mathbb{R}$.
Performing the change of integration variable $x = vt$, we have
\begin{align} \label{eq:spectrum_ode_app}
     \Lambda(k) = \lim\limits_{t\to\infty} \frac{1}{t} \log  \left| 
     \int \rmd v \, \rme^{-\rmi k v t} \exp\left[ \lambda \left(v\right) t \right]
    \right|.
\end{align}
At late times, the integral above can be computed via a saddle-point method. Saddle points, if they exist, solve $\lambda'(v) = \rmi k$.
At $k=0$, this corresponds to a saddle at $v=0$, giving $\Lambda(k=0) = \lambda_0$. At leading order in $k > 0$, the saddle is determined by the leading terms of the small-velocity expansion of $\lambda(v)$
\begin{align} \label{eq:expansion_app}
    \lambda(v \simeq 0) = \lambda_0 \, [1 - (v/v_0)^2] + O(v^4),
\end{align}
with $v_0 > 0$.
This yields complex saddles located at $v = v_0 \left( \frac{\pm\rmi k v_0}{2 \lambda_0} \right) + O(k^2) $. After standard algebra, this finally yields
\begin{align} \label{eq:spectrum_app}
\Lambda(k \simeq 0) &= \lambda_0 - \lambda_0 \left(\frac{v_0 k}{2\lambda_0} \right)^2 + \ldots,
\end{align}
where the ellipsis denotes higher-order terms in $k$ arising from subleading contributions in Eq.~(\ref{eq:expansion_app}).


\paragraph*{(ii) Lyapunov spectrum under stochastic resetting.}
We now apply this formalism to the case of dynamics subject to stochastic resetting. \textit{Mutatis mutandis}, the Lyapunov spectrum is expressed via the generalization of Eq.~(\ref{eq:spectrum_ode_app}) as 
\begin{align} \label{eq:tilde_lambda_k}
   \tilde  \Lambda(k) = \lim\limits_{t\to\infty} \frac{1}{t} \log  \left| 
     \int \rmd v \, \rme^{-\rmi k v t} \exp\left[ \tilde \lambda \left(v\right)  t \right]
    \right|,
\end{align}
where $\tilde \lambda(v)$ is the renormalized velocity-dependent Lyapunov computed in Sect.~\ref{app:lambda_v}.

In the case $r < r_\rmc = \lambda_0$, where  $\tilde \lambda(v) = \lambda(v) - r$, Eq.~(\ref{eq:tilde_lambda_k}) readily yields the renormalized spectrum
\begin{align}
   \tilde  \Lambda(k) =\Lambda(k) - r
\end{align}
for all $k$, where $\Lambda(k)$ is the spectrum in the absence of stochastic resetting expressed in Eq.~(\ref{eq:spectrum_ode_app}).

In the case $r \geq r_\rmc $, the small-velocity expansion of $\tilde \lambda(v)$ reads
\begin{align}
    \tilde \lambda(v) =  \lambda'(v^*) |v| \mbox{ for } v<v^*,
\end{align}
with $\lambda'(v^*) < 0$.
This yields the following integral
\begin{align}
     \int \rmd v \, \rme^{-\rmi k v t} \exp\left[ \lambda'(v^*) |v|  t \right] = - \frac2t \frac{\lambda'(v^*)}{k^2+\lambda'(v^*)^2},
\end{align}
leading to a degenerate vanishing Lyapunov spectrum
\begin{align}
   \tilde  \Lambda(k) = 0
\end{align}
for all $k$.

\section{Case of a parabolic VDLE}
\label{app:parabolic}
Let us consider a parabolic velocity-dependent Lyapunov of the form
\begin{align} \label{eq:model_parabolic}
    \lambda(v) = \lambda_0 \left[1 - \left(\frac{v}{v_\rmB}\right)^2 \right],
\end{align}
where $\lambda_0 >0$ and $v_\rmB$ is the butterfly velocity. This parabolic form is amenable to explicit computations and, notably, the instantaneous eigenvalues of $\tilde D(x,x';t)$ can be determined explicitly, notably without the saddle-point approximations that were used in App.~\ref{app:lambda_v} and App.~\ref{app:lambda_k}.
 
If $r \leq r_\rmc = \lambda_0$, the renormalized VDLE is found to be
\begin{align} \label{eq:tilde_lambda_parabolic}
\tilde \lambda(v) =  \lambda_0 \left[1 - \left(\frac{v}{v_\rmB}\right)^2 \right] - r,
\end{align}
with a corresponding renormalized butterfly velocity $\tilde v_\rmB := \tilde\lambda^{-1}(0) = v_\rmB \sqrt{(r_\rmc -r)/\lambda_0}$, 
and the Lyapunov spectrum reads
\begin{align} \label{eq:tilde_spectrum_parabolic}
    \tilde \Lambda(k) = \lambda_0 \left[ 1  - \left(\frac{v_\rmB} {2\lambda_0} k\right)^2 \right] - r
\end{align}
for all $k \in \mathbb{R}$.

If $r > r_\rmc$: $\tilde \lambda(v)$ develops a linear branch that reads
\begin{align}
    \tilde \lambda(v < v^*) = -2 \frac{v}{v_\rmB} \sqrt{\lambda_0(r-r_\rmc)} 
\end{align}
up to 
\begin{align}
    v^* = v_\rmB \sqrt{\frac{r-r_\rmc}{\lambda_0}}.
\end{align}
The expression of $\tilde \lambda(v\geq v^*)$ is simply given by Eq.~(\ref{eq:tilde_lambda_parabolic}) and $\tilde v_\rmB = 0$.
The Lyapunov spectrum is computed using Eq.~(\ref{eq:spectrum_from_VDLE}),
\begin{align}
        \tilde \Lambda(k) = & \lim\limits_{t\to\infty} \frac{1}{t} \log \left| 
    \int_0^{v^*} \!\! \rmd v \, \cos(kvt) \, \rme^{-2 \frac{v}{v_\rmB} \sqrt{\lambda_0(r-r_\rmc)} t} \right.\nonumber \\
&\quad +\left.   \int_{v^*}^\infty\!\! \rmd v \, \cos(kvt) \, \rme^{ \left(\lambda_0 \left[ 1-(v/v_\rmB)^2\right] - r \right)t}
    \right|.
\end{align}
The first integral above decays algebraically in $1/t$, while the second decays exponentially as $\exp[-2(r-r_\rmc)t]$. Finally, this yields the Lyapunov spectrum
\begin{align}
    \tilde \Lambda(k) =0
\end{align}
for all $k \in \mathbb{R}$.

If $r> r_\rmc$, the renormalized OTOC converges to a steady state $\tilde D_{\rm st}(x,x'):=\lim\limits_{t\to\infty} \tilde D(x,x';t)$ which can be computed directly from the renewal formula:
\begin{align}
    \tilde D_{\rm st}(x,x') = r \int_0^\infty \rmd \tau \, \rme^{\left[\lambda\left(\frac{x-x'}{\tau}\right)-r\right]\tau}.
\end{align}
This yields
\begin{align}
\tilde D_{\rm st}(x,x') = \frac{r}{r-r_\rmc}
     \frac{|x-x'|}{\xi_r}\ K_1\left( |x-x'|/\xi_r \right),
   \end{align}
where $K_1(x)$ is a modified Bessel function of the second kind. At large distances, using $K_1(x) \sim \sqrt{\pi/2x} \exp(-x)$, the expression above describes an OTOC that is exponentially localized around the perturbation site with a localization length $\xi_r = v_\rmB/\sqrt{4\lambda_0 (r-r_\rmc)}$ that diverges at the dynamical transition with an exponent $\nu = 1/2$.
At late times, the approach to the steady state follows
\begin{align}
     \tilde D_{\rm st}(x,x')  - \tilde D(x,x';t) \simeq \rme^{(\lambda_0-r)t} \frac{\lambda_0}{r - \lambda_0}.
\end{align}
This exhibits a timescale $\tau_r := 1/(r-r_\rmc)$ that also diverges at the dynamical transition. Altogether, this critical phenomenon is controlled by a dynamical exponent $z=2$.


\section{Case of a discrete chain}
\label{app:discrete}
The case of a parabolic VDLE can also be explicitly solved on an infinite one-dimensional chain. We consider the parabolic velocity-dependent Lyapunov given in Eq.~(\ref{eq:model_parabolic}) and set the lattice spacing to unity.
The eigenvalues of the OTOC are obtained by discrete Fourier transforms, yielding
\begin{align}
    D(k;t) =
    \rme^{\lambda_0 t} \, \vartheta \left( \frac{k}{2},\exp \left(-\frac{\lambda_0}{v_\rmB^2 t}\right)\right),
\end{align}
 where $\vartheta$ is the Jacobi theta function. The late-time asymptotics yield 
 \begin{align}
   D(k;t) \sim 
\sqrt{\pi \frac{v_{\rmB}^{2} t}{\lambda_{0}}}\;
\rme^{\lambda_0 \left[1
- \left(\frac{v_{\rmB}}{2\lambda_{0}} k\right)^2 \right] t },
 \end{align}
 and we recover the same expression for the Lyapunov spectrum as in the continuum case:
\begin{align}
     \Lambda(k) = \lambda_0 \left[ 1  - \left(\frac{v_\rmB} {2\lambda_0} k\right)^2 \right],
\end{align}
but where $k\in [-\pi, \pi)$.
Interestingly enough, in this discrete case, the condition $\lambda_0 \geq r+ \pi^2/ |2 \lambda_0''|$ guarantees that the Lyapunov spectrum is entirely non-negative.

\bibliography{references}

\end{document}